**Beware of so-called 'good' correlations: a statistical reality check on individual mRNA–protein predictions**

Romain-Daniel Gosselin, PhD

Precision Medicine Unit, Biomedical Data Science Center (BDSC), Lausanne University Hospital (CHUV), Chemin des Roches 1a/1b CH-1010 Lausanne, Switzerland

Tel: +4121 314 2420

Email: Romain-Daniel.Gosselin@chuv.ch

**Subtitle**

Statistical caveats in correlations between RNA and protein contents

**Acknowledgement**

The author thanks Prof. Jacques Fellay for his support.

**Conflict of interest**

The author declares no conflict of interest.

**Summary**

Research in the life sciences often employs messenger ribonucleic acids (mRNA) quantification as a standalone approach for functional analysis. However, although the correlation between the measured levels of mRNA and proteins is positive, correlation coefficients observed empirically are incomplete, necessitating caution in making agnostic inferences. This essay provides a statistical reflection and caveat on the concept of correlation strength in the context of transcriptomics-proteomics studies. It highlights the variability in possible protein levels at given empirical correlation values, even for precise mRNA amount, and underscores the notable proportion of mRNA–protein pairs with abundances at opposite ends of their respective distributions. Cell biologists, data scientists, and biostatisticians should recognise that mRNA–protein correlation alone is insufficient to justify using a single mRNA quantification to infer the amount or function of its corresponding protein.

**The deceiving common sense around a 'good' mRNA–protein correlation**

The quantifications of messenger ribonucleic acids (mRNA) and proteins are omnipresent in life sciences, whether in mechanistic studies or as biomarkers for traits or diseases. The contents of both families of biomolecules are dynamic, providing insight into cell differentiation, physiological pathways, and ongoing pathological processes. The central dogma of molecular biology, which describes the unidirectional flow of information from DNA to mRNA and from mRNA to proteins, fosters an unspoken but misleading model of a proportional relationship between mRNA and matching protein levels. Quantitatively, the observed correlation coefficient between mRNA and protein levels typically hovers around 0.5, whether measured across genes at a genome-wide scale (e.g., for all genes in a specific condition, treatment, tissue, or developmental stage) or across treatments (e.g., for one gene across various conditions) (1-8). The correlation is positive, indicating a global covariation of mRNA and their corresponding proteins, and motivating molecular investigations that often focus solely on mRNA quantification, even when the research concerns gene (i.e., protein) function. However, the observed correlation values merely indicate that as mRNA or protein levels change in a specific direction, the corresponding level of the other molecule is expected to change in the same direction *on average*. This represents the concept of covariance, with correlation being its standardised version. Crucially, the limited strength of the correlation coefficient necessitates caution in interpreting a single mRNA measurement as a surrogate for its corresponding protein, as an imperfect correlation inevitably implies variability, with many mRNA-protein pairs showing abundant mRNA but scarce protein, or *vice versa*.

Certainly, natural variation and measurement error contribute to the scatter and noise in mRNA–protein disagreement to some extent, but the inconsistent regulation of mRNA and their corresponding proteins is a well-documented phenomenon. As Vélez-Bermúdez and Schmidt (9) stated, 'transcript/protein discordance is largely of biological origin (…) and represents a critical layer of regulatory processes at the post-transcriptional level'. These processes include mRNA stability, subcellular protein addressing, protein storage, and degradation, which have been extensively discussed from a biochemical perspective by others (2, 10). This essay, however, takes a statistical approach to the issue. It aims to demonstrate that the concept of an overall 'good correlation' between mRNA and proteins, while possibly defensible depending on the definition of 'good', is fundamentally misleading and cannot be applied agnostically to individual genes. Using mRNA–protein relationships as an example, the essay opportunistically provides a statistical caveat regarding the interpretation of correlation studies. It draws on analytical methods and computer simulations—undeniably oversimplified representations of the actual relationship between mRNA and proteins—but is valid for the purpose of illustration. The narrative, data, and references primarily focus on human and animal research, though the discussion is equally relevant to microbiology and plant science (9, 11).

**A careful look at what a 'good' correlation entails: be wary of ready-made scales**

The concept of statistical correlation pertains to the direction and strength of the overall relationship between two variables. It is mathematised by a coefficient, a single value ranging from −1 to 1. Unless the correlation coefficient is exactly 1 or −1, the relationship between the variables is inherently noisy and exhibits individual variation around the overall trend. Different coefficients exist, with the Pearson's and Spearman's coefficients being the most commonly used. Spearman's coefficient is often preferred because it relaxes many assumptions, such as linearity, required by Pearson's coefficient (see introductory reviews by Schober et al. (12) for a non-mathematical overview). What constitutes a strong correlation? There is no singular, universal answer to this question, but scales are frequently employed to define boundaries of magnitude. Terms such as strong, moderate, or weak, based on cut-off values of the correlation coefficient, are often used to describe the effect size narratively. However, these scales should be interpreted with caution. Each research field involves variables with specific degrees of natural dispersion, which influence the likelihood of different correlation coefficients. Consequently, whether a certain value is deemed a strong correlation depends on normative standards of noise and imprecision within each scientific community. For instance, in social sciences, where large variances are common, correlations above 0.5 are exceptional and regarded as strong (13), whereas the same value might be considered only moderate in biomedical sciences. By definition, the correlation coefficient moves further from 0 (closer to −1 or 1) as data points align more closely with the line assumed to describe the genuine relationship between variables (e.g., a straight line for Pearson's correlation). In other words, the correlation coefficient corresponds to the absence of dispersion in the covariation between variables, which determines the predictability of one variable based on the other. Thus, one way to address the elusive concept of correlation strength is to use a tangible analytical measure of this dispersion. The next section explores its mathematical meaning. Although not overly advanced in mathematics for statistician readers—given both the limited utility of an in-depth demonstration and the author's admitted mathematical limitations—this chapter may feel daunting or unnecessary for non-statisticians. If so, readers are encouraged to skip ahead to the following chapter titled *General reflection on the limitations of a positive correlation*, which offers a narrative discussion of the analytical section below.

**Analytically analysing the dispersion in a positively correlated normal bivariate space**

Consider a transcriptomics-proteomics normal bivariate distribution with a specified Pearson's correlation coefficient (Figure 1A). While this distributional approximation is undeniably an oversimplification for the relationship between mRNA and proteins, it facilitates the analytical calculation of probabilities within specific intervals. The conditional probability of a protein level, given a particular mRNA level, follows a normal distribution as described below:

$$(protein|mRNA = x) \sim \mathcal{N}\left(\mu_{protein} + \frac{\rho \sigma_{proetein}(x - \mu_{mRNA})}{\sigma_{mRNA}}, (1 - \rho^2)\sigma^2_{protein}\right)$$

where $\mu_{mRNA}$ and $\sigma_{mRNA}$ are the mean and standard deviation of the mRNA content, $\mu_{protein}$ and $\sigma_{protein}$ are the mean and standard deviation of the protein content, and $\rho$ is the mRNA–protein Pearson's correlation coefficient. Thus, the probability of a protein level conditional to a particular mRNA level has a mean of $\mu_{protein} + \frac{\rho \sigma_{proetein}(x - \mu_{mRNA})}{\sigma_{mRNA}}$ and a variance of $(1 - \rho^2)\sigma^2_{protein}$.

We may then pose the following question: What is the probability that protein content is above average if mRNA content is exactly one standard deviation below average? To answer this, the first step is to set the correlation coefficient and the marginal parameters for each variable. These values are derived from the empirical dataset obtained from the atlas established by Wang et al. (7), calculated by averaging the summary statistics from five tissues (adrenal gland, appendix, brain, colon and duodenum). The parameters ($Log_{10}$) used are as follows: mean mRNA expression (measured in fragments per kilobase of transcript per million mapped reads [FPKM]) = 1.0, standard deviation of mRNA = 0.5, mean protein expression (measured in intensities based on the area under the curve [IBAQ]) = 6.9, standard deviation of protein = 1.1, and Pearson's correlation coefficient = 0.5. Substituting these terms into the equation, we obtain:

$$(protein|mRNA = 0.5) \sim \mathcal{N}\left(6.9 + \frac{(0.5)(1.1)(0.5 - 1.0)}{0.5}, (1 - 0.5^2)1.1^2\right)$$

$$(protein|mRNA = 0.5) \sim \mathcal{N}(6.35, 0.90)$$

This indicates that the protein level, conditional on an mRNA concentration of 0.5 FPKM, follows a normal distribution with a mean of 6.35 IBAQ and a variance of 0.90 IBAQ (corresponding to a standard deviation of 0.95). Based on this, we can determine the probability that the protein content exceeds 6.9 IBAQ at the exact mRNA value of 0.5. To do this, we calculate the Z-score:

$$Z = \frac{6.9 - 6.35}{0.95} = 0.58$$

This corresponds to the following probability:

$$p(Z \geq 0.65) = 0.28$$

Therefore, with an mRNA-protein correlation coefficient of 0.5, a sizeable proportion (28%) of proteins exhibit levels above average, even when their corresponding mRNA level is one standard deviation below average. This is illustrated in Figure 1B, which depicts a simulated sample of 5000 data points.

Extending the analysis further, we may ask: What is the probability that the protein content is *at least* the average while the mRNA content is *at most* the average? This was calculated using the cumulative density function of the bivariate normal distribution. The Statistics Online Computational Resource (SOCR) public simulator, provided by the University of Michigan School of Nursing and accessible at https://socr.umich.edu/HTML5/BivariateNormal/, was used for this purpose. The results showed that proportion of genes with protein content higher than the average (6.9) while mRNA content remains below the average (1.0) is 0.17. Symmetrically, the probability of protein levels being lower than average while mRNA content is higher than average is also 0.17. With the number of mRNA–protein pairs rounded to 15,000, this translates to approximately 5,100 (34%) genes with protein and mRNA levels on opposite sides of their respective averages. This is illustrated in Figure 1C using simulated data.

**General reflection on the limitations of a positive correlation**

In the previous section, using a bivariate normal model with realistic parameter values, we presented two objective calculations that contribute to the discussion of the implications of a 0.5 correlation. First, one-fourth to one-third of mRNA with abundances of one standard deviation below the mean are expected to have corresponding protein levels exceeding the mean protein content. Second, nearly one-third of genes are expected to exhibit abundant mRNA and scarce protein, or *vice versa*.

Whether these numerical results are sufficient to challenge and reconsider the narrative of a 'strong' correlation is subjective and ultimately left to the reader's judgement. However, they clearly demonstrate that labelling a 0.5 correlation as strong and using it to infer a correspondence between mRNA and protein levels for a specific gene is a bold assumption. Some readers may point out that certain studies have reported higher mRNA–protein correlation coefficients. Yet, as shown in Table 1, which extends the above calculations across Pearson's correlation values ranging from 0.1 to 0.9, even at higher correlation coefficients, significant discrepancies between mRNA and protein levels persist. For instance, a correlation of 0.8 results in approximately one-tenth (9%) of mRNA levels one standard deviation below average corresponding to protein levels above average, and nearly one-fifth (21%) of genes exhibiting an inverse relationship between mRNA and protein abundance. Thus, the caution against relying on a single mRNA level to predict its corresponding protein content remains valid, even when the correlation is optimistically set at a higher value.

| Pearson's r | Mean [protein] given [mRNA] = 0.50 | Standard deviation of [protein] given [mRNA] = 0.50 | Z score of [protein] ≥ 6.90 | Probability to get at least Z (upper probability) | Probability of mismatch (from SOCR calculator) |
|---|---|---|---|---|---|
| 0.1 | 6.8 | 1.1 | 0.10 | 0.46 | 0.46 |
| 0.2 | 6.7 | 1.1 | 0.20 | 0.42 | 0.44 |
| 0.3 | 6.6 | 1.0 | 0.31 | 0.38 | 0.40 |
| 0.4 | 6.5 | 1.0 | 0.44 | 0.33 | 0.36 |
| 0.5 | 6.3 | 1.0 | 0.58 | 0.28 | 0.34 |
| 0.6 | 6.2 | 0.9 | 0.75 | 0.23 | 0.29 |
| 0.7 | 6.1 | 0.8 | 0.97 | 0.16 | 0.26 |
| 0.8 | 6.0 | 0.7 | 1.33 | 0.09 | 0.21 |
| 0.9 | 5.9 | 0.5 | 2.06 | 0.02 | 0.14 |

*Table 1: Influence of mRNA–protein correlation coefficient (Pearson) on the mean and standard deviation of protein levels for mRNA is fixed at 0.5 FPKM (1 standard deviation below average), Z score and corresponding upper probability for protein at least average, and probability of mismatch.*

**Estimation of mismatching mRNA–protein pairs by Monte Carlo simulation**

Protein and mRNA contents are asymmetrically bounded at either zero because concentrations cannot be negative or at a specified cut-off value, such as in the study by Wang *et al.* where the authors used a cut-off FPKM of 1 ($\log_{10}$ [FPKM]= 0). Consequently, their distributions are slightly truncated with a longer tail toward higher concentrations and should ideally be simulated as such. The previous analytical approach using a bivariate normal distribution only partially reflects this reality. To address this, we simulated a truncated bivariate normal distribution using the same parameters (from the empirical dataset) as in the normal bivariate scenario (7). Figure 2A illustrates the similarity between the resulting family of distributions and the actual mRNA and protein quantifications from the study by Wang et al. (7). A Monte Carlo simulation was then employed to generate 5000 iterations of random samples from this truncated bivariate normal distribution, yielding point and interval estimates of the proportion of genes with mRNA and protein levels in opposite directions (i.e. mismatches, shown in the red quadrants in Figure 2B). As depicted in Figure 2C, which displays the sampling distribution of mismatches, our simulations produced a median count of 5,098 genes out of 15,000 (interquartile range [5,060–5,138], range [4,898–5,298]) with mRNA and protein levels of opposite abundances. This fraction closely aligns with the value previously calculated analytically using the untruncated bivariate normal model, indicating that the truncation has minimal impact on the occurrence of mismatches.

**Achieving better correlation and making the most of mRNA investigation**

It is important to emphasise that this essay does not seek to discredit or dismiss mRNA investigations. In the case of the aforementioned across-gene studies, the correlations correspond to genome-wide quantifications, encompassing the entire transcriptome and proteome. Higher mRNA–protein correlations have been reported when focusing on genes whose mRNA regulation is *a priori* known or expected to translate into protein and functional changes, such as those involved in the cell cycle, which are notably time-regulated (14). Similarly, increased correlations have been observed across conditions when investigations concentrate on mRNAs identified *a priori* as differentially expressed between treatments (15).

Other examples of improved correlations include studies that restricted quantifications to ribosome-bound nucleic acids (7), focused on genes within accessible chromatin regions (16), applied corrective mathematical approaches for prediction (17), or employed machine learning algorithms utilising the levels of multiple transcripts for single predictions (18).

**Conclusions and limitations**

In this study, using parameters derived from empirical observations, we applied a statistical approach to question the agnostic use of mRNA levels as a proxy for protein abundance. At typical mRNA–protein correlation levels, the substantial proportion of genes exhibiting opposing enrichments of these molecules underscores the need for careful interpretation when deducing protein content from mRNA quantifications. As life sciences increasingly integrate large-scale omics projects, fostering a culture of well-understood statistical standards becomes ever more essential. While the data generated by these projects undoubtedly hold immense potential for advancing our understanding of biological systems, failing to contextualise mRNA content with protein levels risks overgeneralising conclusions, potentially leading to biased biological models and misinformed decisions.

The empirical parameters were selected or calculated based on only five datasets from the transcriptome and proteome atlas published by Wang et al. (7). While the datasets from the numerous tissues included in this study exhibit remarkable consistency both among themselves and with most available literature, certain experimental settings or cell types might display different properties. However, as demonstrated above, such differences would affect the quantitative but not the qualitative nature of the caution associated with inferring single protein levels from mRNA, even at higher correlation levels.

**Methods**

Statistical analyses and simulations were conducted using RStudio version 2024.09.1+394 (RStudio Team (2020). RStudio: Integrated Development for R. RStudio, PBC, Boston, MA. URL: http://www.rstudio.com/). Graphs were also created using RStudio. To characterise the bivariate normal distributions, contour plots were generated using the MASS package, with empirical parameters derived from the human transcriptome proteome atlas by Wang et al. (7). The summary statistics (mean and standard deviation) were calculated from raw data of 5 tissues (adrenal gland, appendix, brain, colon and duodenum) downloaded from supplementary Tables EV1 and EV2 of the Wang study and averaged (see supplementary material and the Readme tab on the Figshare repository at https://doi.org/10.6084/m9.figshare.28138733.v1). Similarly, the correlation coefficient used (0.5) is the average of values from 5 tissues taken directly from Appendix Figure S11 of this same article. Random samples from these distributions were drawn using the rmvnorm package. Truncated bivariate normal distributions and Monte Carlo simulations (n = 5,000 iterations of n = 15,000 each) were generated using the tmvtnsim and truncnorm packages. The R code was drafted by the author and debugged, when necessary, using the Claude chatbot (https://claude.ai) provided by Anthropic, PBC. The simulation code, organised by matching figure, as well as the Excel file used to calculate summary statistics, is freely available on the Figshare repository (https://doi.org/10.6084/m9.figshare.28138733.v1) to ensure reproducibility. Throughout the text, proportions, probabilities and Z scores were rounded to two decimal places, summary statistics to the first decimal place, and percentages were rounded to the nearest integer.

# References

1. Bosi C, Bartha A, Galbardi B, Notini G, Naldini MM, Licata L, et al. Pan-cancer analysis of antibody-drug conjugate targets and putative predictors of treatment response. Eur J Cancer. 2023;195:113379.
2. Buccitelli C, Selbach M. mRNAs, proteins and the emerging principles of gene expression control. Nat Rev Genet. 2020;21(10):630-44.
3. Gry M, Rimini R, Stromberg S, Asplund A, Ponten F, Uhlen M, Nilsson P. Correlations between RNA and protein expression profiles in 23 human cell lines. BMC Genomics. 2009;10:365.
4. Maier T, Guell M, Serrano L. Correlation of mRNA and protein in complex biological samples. FEBS Lett. 2009;583(24):3966-73.
5. Moritz CP, Muhlhaus T, Tenzer S, Schulenborg T, Friauf E. Poor transcript-protein correlation in the brain: negatively correlating gene products reveal neuronal polarity as a potential cause. J Neurochem. 2019;149(5):582-604.
6. Takemon Y, Chick JM, Gerdes Gyuricza I, Skelly DA, Devuyst O, Gygi SP, et al. Proteomic and transcriptomic profiling reveal different aspects of aging in the kidney. Elife. 2021;10.
7. Wang D, Eraslan B, Wieland T, Hallstrom B, Hopf T, Zolg DP, et al. A deep proteome and transcriptome abundance atlas of 29 healthy human tissues. Mol Syst Biol. 2019;15(2):e8503.
8. Wegler C, Olander M, Wisniewski JR, Lundquist P, Zettl K, Asberg A, et al. Global variability analysis of mRNA and protein concentrations across and within human tissues. NAR Genom Bioinform. 2020;2(1):lqz010.
9. Velez-Bermudez IC, Schmidt W. The conundrum of discordant protein and mRNA expression. Are plants special? Front Plant Sci. 2014;5:619.
10. Liu Y, Beyer A, Aebersold R. On the Dependency of Cellular Protein Levels on mRNA Abundance. Cell. 2016;165(3):535-50.
11. Ponnala L, Wang Y, Sun Q, van Wijk KJ. Correlation of mRNA and protein abundance in the developing maize leaf. Plant J. 2014;78(3):424-40.
12. Schober P, Boer C, Schwarte LA. Correlation Coefficients: Appropriate Use and Interpretation. Anesth Analg. 2018;126(5):1763-8.
13. Hemphill JF. Interpreting the magnitudes of correlation coefficients. Am Psychol. 2003;58(1):78-9.
14. Ly T, Ahmad Y, Shlien A, Soroka D, Mills A, Emanuele MJ, et al. A proteomic chronology of gene expression through the cell cycle in human myeloid leukemia cells. Elife. 2014;3:e01630.
15. Koussounadis A, Langdon SP, Um IH, Harrison DJ, Smith VA. Relationship between differentially expressed mRNA and mRNA-protein correlations in a xenograft model system. Sci Rep. 2015;5:10775.
16. Sanghi A, Gruber JJ, Metwally A, Jiang L, Reynolds W, Sunwoo J, et al. Chromatin accessibility associates with protein-RNA correlation in human cancer. Nat Commun. 2021;12(1):5732.
17. Edfors F, Danielsson F, Hallstrom BM, Kall L, Lundberg E, Ponten F, et al. Gene-specific correlation of RNA and protein levels in human cells and tissues. Mol Syst Biol. 2016;12(10):883.
18. Prabahar A, Zamora R, Barclay D, Yin J, Ramamoorthy M, Bagheri A, et al. Unraveling the complex relationship between mRNA and protein abundances: a machine learning-based approach for imputing protein levels from RNA-seq data. NAR Genom Bioinform. 2024;6(1):lqae019.

**Figure Legends**

**Figure 1: Analytical analysis of mRNA and protein scatter and mismatch**

(A) Contour plot illustrating the theoretical mRNA–protein bivariate normal distribution with a correlation coefficient of 0.5. The concentric ellipses represent lines of constant data density, indicating the amplitude of the third dimension of the graph. The normal distributions in the x (mRNA) and y (protein) dimensions are shown in the two marginal distributions. Empirical parameters were extracted from Wang et al. (7). Grey lines at x = 1.0 and y = 6.9 indicate the mean values. (B) Scatter plot of a simulated (n = 5,000) mRNA–protein dataset, highlighting the proportion of data (red) with protein levels above the average (IBAQ > 6.9, red) among those with mRNA levels one standard deviation below the average (FPKM = 0.5, blue). The one standard deviation value is shown as a blue dotted line. For visibility and to avoid increasing the sample size, FPKM values between 0.48 (0.96 standard deviations) and 0.52 (1.04 standard deviations) were included in the data displayed in blue and red. (C) The same scatter plot as in B, showing data points with mismatching mRNA and protein levels (red). The expected proportion, calculated analytically, is 0.34.

**Figure 2: Simulation of a truncated bivariate correlation**

(A) Distributions of mRNA (dark grey) and protein (blue) abundances in simulated truncated bivariate normal data (n = 15,000). The distributions closely resemble those empirically reported by Wang et al. (7) (insert, example of the brain dataset). (B) Example scatter plot of a truncated bivariate normal dataset, with mismatching data points shown in red. (C) Sampling distribution of mismatch counts obtained through Monte Carlo simulation (n = 5,000 iterations of n = 15,000 observations).

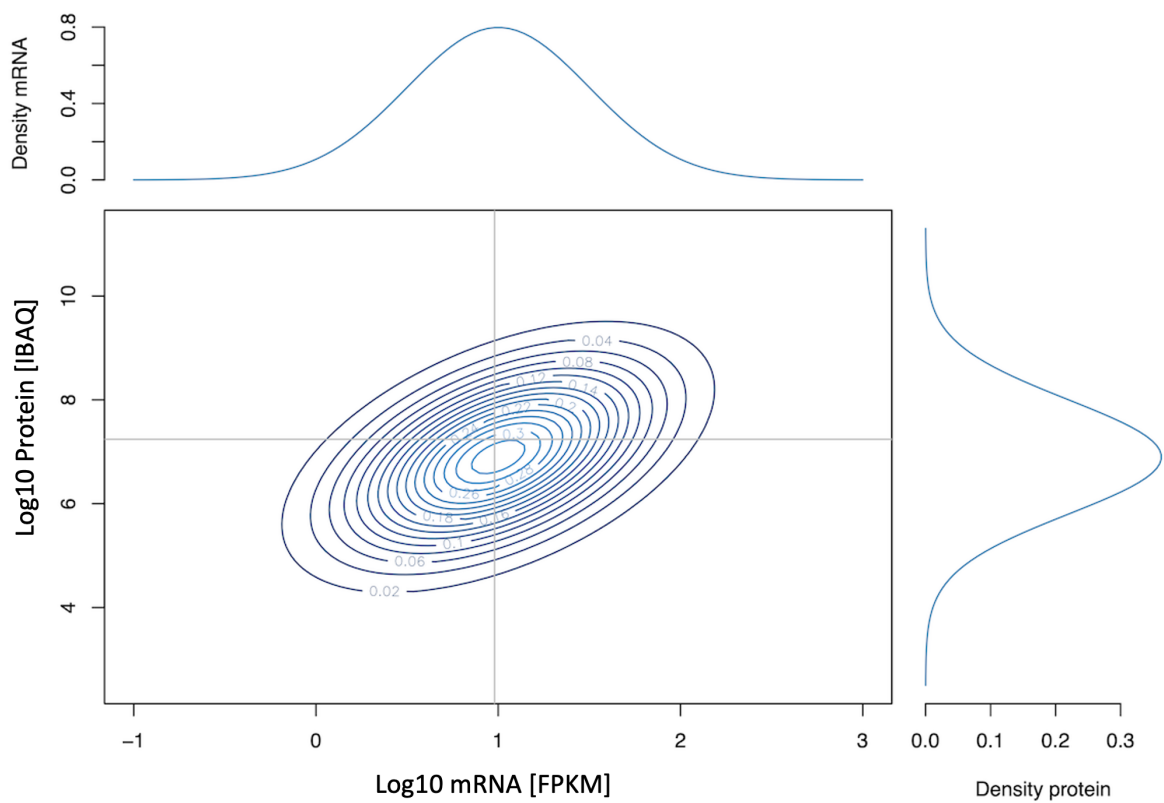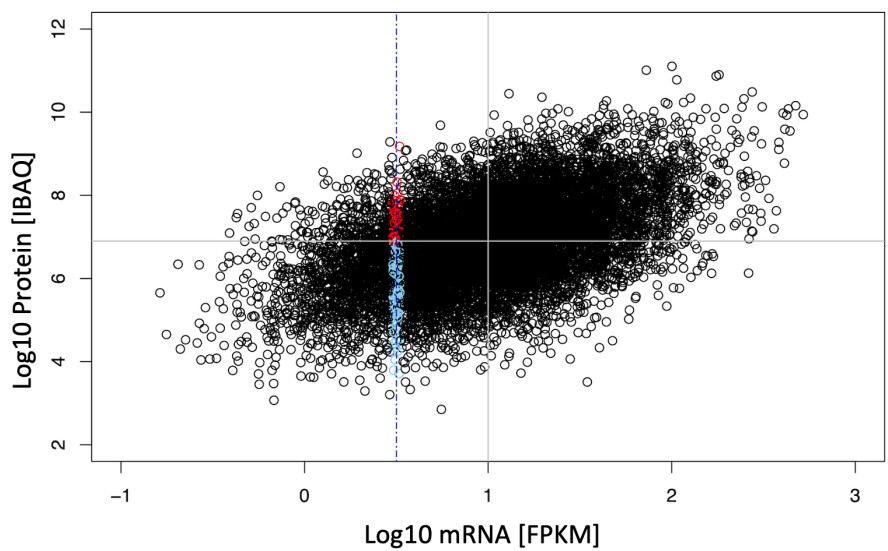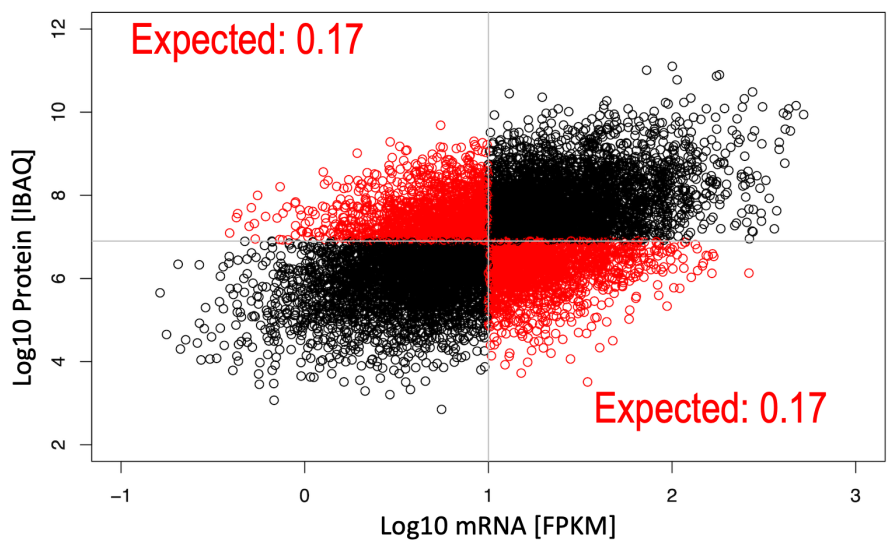

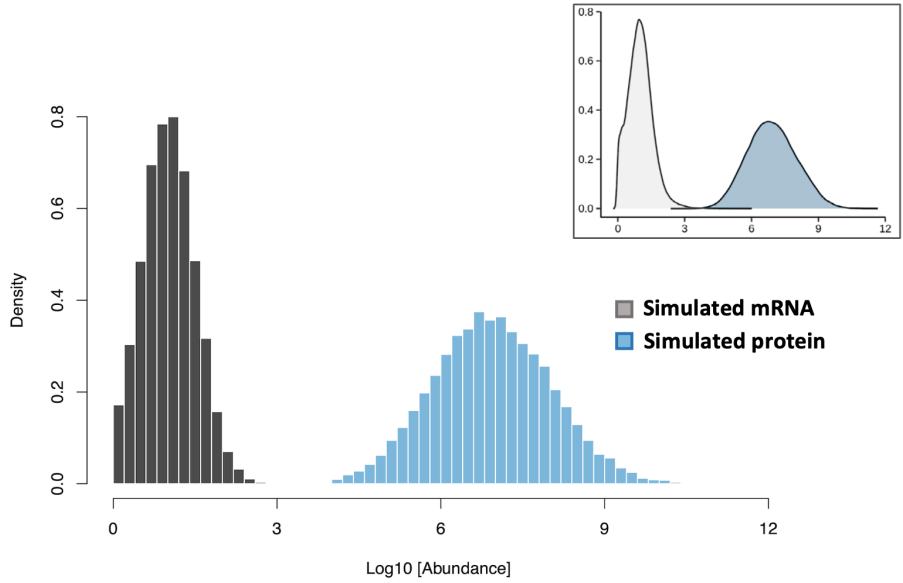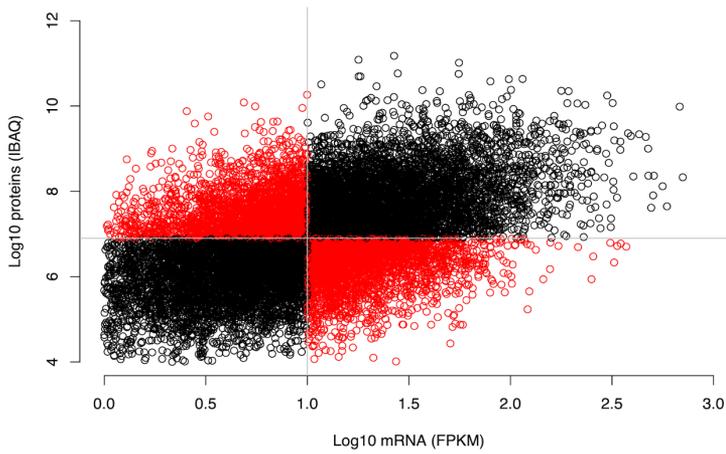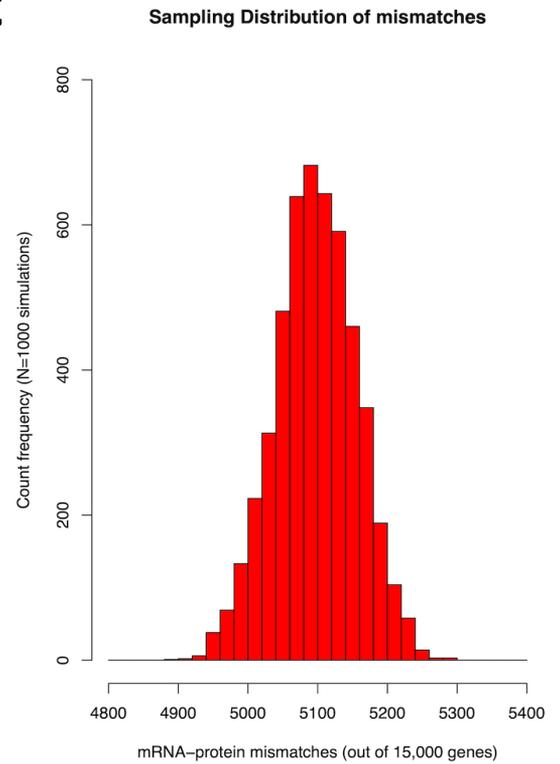